\documentclass{mem}
\usepackage{natbib}\usepackage{txfonts}\usepackage{balance}
\usepackage{graphicx}
\usepackage[a4paper,breaklinks,dvipdfm]{hyperref}
\idline{75}{282}
\begin{document}
\def\teff{$T\rm_{eff }$}
\def\kms{$\mathrm {km s}^{-1}$}

\title{
NGC~5694: another extra-galactic globular cluster}


\author{
A. Mucciarelli\inst{1},
M. Bellazzini\inst{2},
M. Catelan\inst{3,4},
P. Amigo\inst{3,4},
M. Correnti\inst{5},
C. Cort\'{e}s\inst{6},\\
E. Dalessandro\inst{1}
\and V. D'Orazi\inst{7.8}}

  \offprints{A.Mucciarelli}

\institute{
Dipartimento di Fisica \& Astronomia, Universit\`a 
  degli Studi di Bologna, Viale Berti Pichat, 6/2 - 40127, 
  Bologna, Italy
\and
INAF-Osservatorio Astronomico di Bologna, Via Ranzani, 1 - 40127, 
  Bologna, Italy
  \and
  Pontificia Universidad Cat\'olica de Chile, Facultad de F\'isica, 
  Departamento de Astronom\'ia y Astrof\'isica, Av. Vicu\~na Mackena 4860, 
  782-0436 Macul, Santiago, Chile
   \and
  The Milky Way Millennium Nucleus, Santiago, Chile 
  \and
  INAF-Istituto di Astrofisica Spaziale e Fisica Cosmica, Via Gobetti 101,
  I-40129, Bologna, Italy
  \and
  Departamento de F\'isica, Facultad de Ciencias B\'asicas, Universidad Metropolitana de Ciencias 
  de la Educaci\'on, Av. Jos\'e Pedro Alessandri 774, 
  776-0197 \~{N}u\~noa, Santiago, Chile
  \and
Department of Physics and Astronomy, Macquarie University, Balaclavard, NorthRyde, NSW2109, Australia
\and
Monash Centre for Astrophysics, School of Mathematical Sciences, Building 28, Monash 
University, VIC3800, Australia   
}

\authorrunning{Mucciarelli}


\abstract{
We discuss the chemical composition of six giant stars of the outer Halo globular cluster 
NGC~5694, through the analysis of UVES@FLAMES high-resolution spectra. The cluster has 
an average iron content [Fe/H]=--1.83$\pm$0.01, solar-scaled [$\alpha$/Fe] ratios and a
very low Ba abundance ([Ba/Fe]=--0.71$\pm$0.06). These anomalous abundance patterns are 
different from those observed in other Halo globular clusters but similar to those of the 
metal-poor stars 
in typical dwarf spheroidal galaxies. These findings suggest an extra-galactic origin for 
NGC~5694, likely from a dwarf spheroidal galaxy.
}
\maketitle{}

\section{Introduction}

NGC~5694 is an old, metal-poor globular cluster (GC) located in the outer Halo, at a 
galactocentric distance of 30 kpc \citep{lee}. This cluster has not received a great 
deal of attention and only a few of works are devoted to the study of its chemical and kinematical 
properties, pointing out two peculiarities: 
{\sl (i)}~a large radial velocity, $V_{rad}$=--140.7 km/s, as estimated by
\citet{geisler95}, and {\sl (ii)}~a peculiar chemical composition in one giant star 
discussed by \citet{lee}.\\
The unusual abundance pattern of the only star studied so far 
(with solar-scaled abundance ratios for the $\alpha$-elements and sub-solar 
abundance ratios for neutron-capture elements) 
is unique among the halo GC, suggesting a possible extra-galactic origin for this 
star cluster, as suggested for other 
(more metal-rich) GCs with anomalous abundance ratios, as Ruprecht 106 \citep{brown97}, Palomar 12 
\citep{cohen04} and Terzan 7 \citep{sbordone07}.

\section{Chemical analysis}
In order to confirm the hypothesis of an extra-galactic origin for NGC~5694 proposed by 
\citet{lee} and to provide a complete 
description of the chemical composition of this GC,
we analysed high-resolution spectra acquired with the spectrograph UVES@FLAMES (VLT) 
of six giant stars located in the bright portion of the Red Giant Branch of the cluster. 
The targets have been selected among cluster members already confirmed by \citet{geisler95}. 
Atmospheric parameters have been derived 
from the photometry by \citet{correnti}.\\
As a sanity check, we analysed UVES@FLAMES spectra retrieved from the ESO archive of 13 giant stars 
in the Galactic GC NGC~6397, taken as prototype of a genuine Halo cluster with the same 
metallicity of NGC~5694. The analysis of NGC~6397 has been performed by using the same linelist and 
the same methodology adopted for NGC~5694.

\section{Chemical composition of NGC~5694}
\begin{figure}[h]
\resizebox{\hsize}{!}{\includegraphics[width=5.5in]{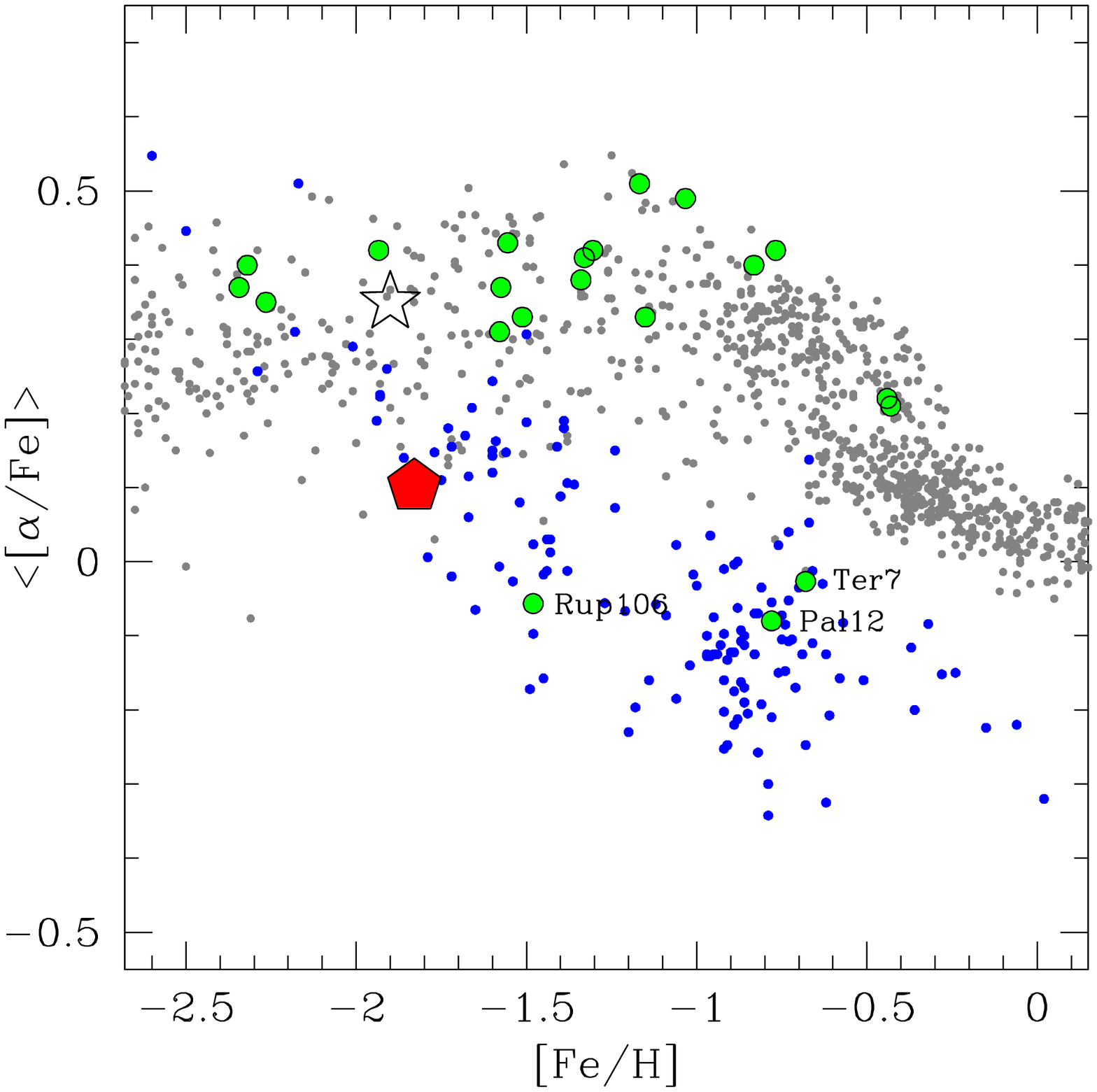}}
\caption{\footnotesize $<$[$\alpha$/Fe]$>$ abundance ratios 
as a function of [Fe/H]: NGC~5694 is marked as a red pentagon. 
Grey points are Galactic field stars \citep{edvardsson93,reddy03,reddy06}; 
green circles are Galactic GCs \citep{brown97,cohen04,sbordone07,carretta09}, 
blue points are stars of dwarf spheroidal galaxies \citep{shetrone01, shetrone03,sbordone07,letarte10,lemasle12}.
NGC~6397 is shown as a empty star symbol. 
}
\label{alfa}
\end{figure}

\begin{figure}[h]
\resizebox{\hsize}{!}{\includegraphics[width=3.5in]{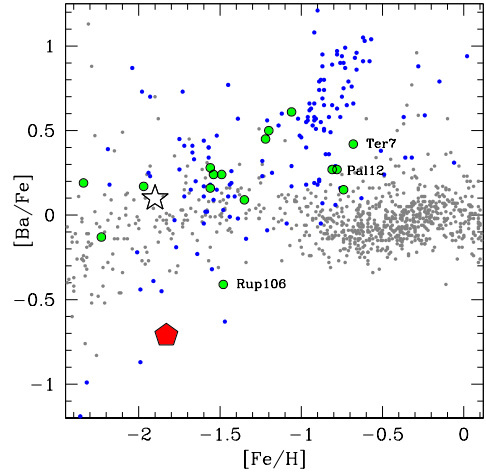}}
\caption{\footnotesize [Ba/Fe] abundance ratios as a function of [Fe/H]. 
Same symbols of Fig.~\ref{alfa}. Data for the Galactic GCs are from \citet{dorazi}.
}
\label{barium}
\end{figure}

\begin{figure}[h]
\resizebox{\hsize}{!}{\includegraphics[width=3.5in]{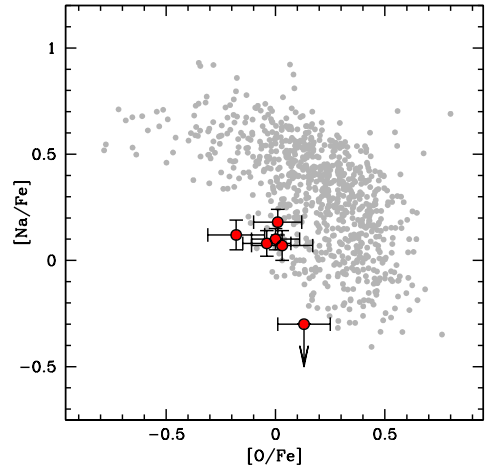}}
\caption{\footnotesize [Na/Fe] abundance ratios as a function of [O/Fe]. 
Red circles are the six individual stars of NGC~5694.
Grey points are individual stars of Galactic GCs by \citet{carretta}.
}
\label{nao}
\end{figure}

The main results concerning the chemical composition of NGC~5694 are summarised as follows:
\begin{itemize}
\item the six stars studied do not show any detectable spread in the abundances of 
Fe, O, Mg, Si, Ca, Ti and Ba. For this reason we will refer to the average abundance of the 
six stars as to the cluster abundance;
\item the cluster has an iron content of [Fe/H]=~--1.83$\pm$0.01  from the analysis 
of single ionised lines. Neutral iron lines provide a lower abundance ([Fe/H]=~--1.97$\pm$0.03) 
because of the occurrence of NLTE effects. No evidences of intrinsic spread in the iron content of the cluster 
have been found. The derived metallicity well agrees with that provided by \citet{lee};
\item NGC~5694 is remarkably deficient in $\alpha$-elements. [O/Fe], [Mg/Fe], [Ca/Fe] and [Ti/Fe] 
abundance ratios are solar-scaled, while only the [Si/Fe] ratio is enhanced with respect to the solar value. 
These values are lower than those usually observed in Galactic GCs of similar metallicity. 
Fig.~\ref{alfa} shows the position of NGC~5694 in the [$\alpha$/Fe] vs [Fe/H] diagram, 
in comparison with Galatic GCs (green points), Halo field stars (grey points) and dwarf spheroidal 
galaxies stars (blue points); the position of NGC~6397 as derived by our differential analysis 
is shown as a star symbol. Also, the position of three anomalous GCs (namely, Ruprecht 106, Palomar 12 and 
Terzan 7) is explicitly shown;
\item NGC~5694 exhibits a peculiar under-abundance of the [Ba/Fe] ratio, with an average value 
[Ba/Fe]=--0.70$\pm$0.06. 
As shown in Fig.~\ref{barium}, [Ba/Fe] of the cluster is $\sim$0.5-0.7 dex lower than in the Milky Way 
stars; in particular, the Halo GCs with metallicities similar to that of NGC~5694 have solar-scaled
[Ba/Fe] ratios \citep{dorazi};
\item Fig.~\ref{nao}  shows the position of the six individual stars of NGC~5694 
in the [Na/Fe]--[O/Fe] plane, in comparison with the individual stars analysed in 19 Galactic GCs 
by \citet{carretta}. 
One star of NGC~5694 out of six exhibit a lower [Na/Fe] abundance ratio with respect to all the others.
Instead, O abundances do not show hints of intrinsic spread. The low Na abundance (coupled with a 
low Al abundance) in this star is compatible with the occurrence of self-enrichment processes in the early stages 
of the cluster life. 
\end{itemize}

\section{Conclusions}
The analysis of six giant stars of NGC~5694 confirms the first results provided by \citet{lee} 
based on a single star.
The cluster shows unusual low [$\alpha$/Fe] abundance ratios, lower than those observed in 
Galactic (both field and GCs) stars of similar metallicity. 
Also, NGC~5604  is remarkably deficient of Ba abundance, with a [Ba/Fe] ratio 
lower then the Galactic stars by $\sim$0.7 dex. 
In light of these findings, NGC~5694 formed from a gas already enriched by Type Ia Supernovae 
and likely in a galactic environment characterised by a star formation rate slower than 
that typical of the Galactic Halo.
These anomalous chemical patterns are incompatible with those observed in other halo globular clusters, 
but they resemble the chemical patterns observed in the nearby dwarf spheroidal galaxies 
\citep[see e.g.][]{tolstoy}. 
These findings suggest an extra-galactic origin for NGC~5604, likely formed in a dwarf galaxies 
and then accreted by the Galactic halo.

%

%
%

\bibliographystyle{aa}

\end{document}